\documentclass{ifacconf}
\usepackage{color}
\usepackage{bm}
\usepackage{epsfig}
\usepackage{pdfpages}
\usepackage{physics}
\usepackage{amssymb,amsbsy}
\usepackage{bbm}
\usepackage{amsmath,tabularx}
\usepackage{enumitem}  
\usepackage{mathrsfs}  
\usepackage{pdfpages}
\usepackage{algorithm}
\usepackage{dsfont}
\usepackage{mathtools}
\mathtoolsset{showonlyrefs}
\usepackage{subcaption}
\usepackage[noend]{algpseudocode}
\usepackage{pdflscape}
\usepackage{arydshln}
\usepackage[UKenglish]{babel}
\usepackage{color}
\usepackage{textcomp}
\usepackage{qtree}
\usepackage{tikz}
\usepackage{tikz-qtree}
\usepackage[xcolor]{changebar}  \cbcolor{red}
\usepackage[acronym]{glossaries}
\usepackage{graphicx}      
\usepackage{natbib}        
\newacronym{MA}{MP}{Max-Plus}
\newacronym{SMPL}{SMPL}{Switching Max-Plus Linear}
\newacronym{MMPS}{MMPS}{Max-Min-Plus-Scaling}
\newacronym{DEHA}{DEHA}{Discrete-Event Hybrid Automata}

\makeatletter
\def\BState{\State\hskip-\ALG@thistlm}
\makeatother

\newcommand{\MA}{max-plus }

\newcommand{\maxpow}[2]{{#1}^{{\scriptscriptstyle\otimes}^{\scriptstyle{#2}}}} 
\newcommand{\RN}[1]{%
	\textbf{\uppercase\expandafter{\romannumeral#1}}%
}
\newcommand{\rn}[1]{%
	\textit{\lowercase\expandafter{\romannumeral#1})}%
}

\def\LRES{{%
		\setbox0\hbox{$\backslash$}%
		\rlap{\hbox to \wd0{\hss$\circ$\hss}}\box0
}}

\def\RRES{{%
		\setbox0\hbox{$/$}%
		\rlap{\hbox to \wd0{\hss$\circ$\hss}}\box0
}}
\newtheorem{definition}{Definition}
\numberwithin{definition}{section}

\newtheorem{theorem}{Theorem}
\numberwithin{theorem}{section}

\numberwithin{lemma}{section}
\newtheorem{proposition}{Proposition}
\numberwithin{proposition}{section}

\numberwithin{corollary}{section}

\numberwithin{remark}{section}

\numberwithin{example}{section}
\theoremstyle{break}
\usepackage{setspace}
\begin{document}
	
	\begin{frontmatter}
		
		\title{Framework for Studying Stability of Switching Max-Plus Linear Systems} 
		
		
		\author[1]{Abhimanyu Gupta} 
		\author[1]{Ton van den Boom}
		\author[2]{Jacob van der Woude} 
		\author[1]{Bart De Schutter}
		
		\address[1]{Delft Center for Systems and Control, Delft University of Technology, Delft, The Netherlands}
		\address[2]{DIAM, EWI, Delft University of Technology, Van Mourik Broekmanweg 6, 2628 XE, Delft, The Netherlands\\{e-mail: \{\ttfamily{a.gupta-3,a.j.j.vandenboom,j.w.vanderwoude,b.deschutter}\}@tudelft.nl}}

		\begin{abstract}:                
			We propose a framework for studying the stability of discrete-event systems modelled as switching max-plus linear systems. In this framework, we propose a set of notions of stability for generic discrete-event systems in the max-plus algebra. Then we show the loss of equivalence of these notions for switching max-plus linear systems due to the lack of global monotonicity and the accompanying difficulty in rigorous analysis. This serves as a motivation to relax the assumption on monotonicity of the dynamics to positive invariance of max-plus cones. Then we proceed to generalise the notions of stability when the dynamics is restricted to such cones. The stability analysis approach presented in this paper serves as a first step to study the stability of a general class of switching max-plus linear systems.
		\end{abstract}
		
		
	\end{frontmatter}
	
	\section{Introduction}
	Max-plus and associated idempotent semiring structures allow the study of synchronisation behaviour and timing aspects of discrete-event systems in a linear fashion \citep{Baccelli1992}. The associated systems theory finds applications in analysis of productions systems, queueing systems, timetabling of transportation networks, and so on \citep{Komenda2018}. Most importantly, it offers a convenient framework to study stability of discrete-event systems \citep{Baccelli1992,Commault1998a}.  

	Stability analysis plays an important role in the operation and control of a dynamical system. The aim of this paper is to introduce a framework for analysing stability of discrete-event systems modelled as a \Gls{SMPL} system. Such models extend the max-plus linear modelling framework by allowing changes in synchronisation and ordering constraints in the system evolution \citep{VandenBoom2006,VanDenBoom2012b}. 
	
	
	There has been considerable research in generalising the notions of stability from linear time-invariant to switching systems in conventional algebra \citep{Liberzon1999}. The research for the counterpart in max-plus algebra is limited. Most of the existing literature is centred around the existence and uniqueness of stationary regimes of the dynamics governed by a class of monotone and additively homogeneous functions in the max-plus algebra \citep{Mairesse1997,Merlet2010a,Katz2012}. The class of \Gls{SMPL} systems is, however, larger than the ones described in \citep{Mairesse1997,Merlet2010a,Katz2012}. An \Gls{SMPL} system can have, in general, multiple stationary regimes that correspond to different asymptotic growth rates of the states. This is due to the loss of global monotonicity of the dynamics. Therefore, the existing sufficient conditions for stability in the literature are restrictive for the purpose of control and analysis.	
	
	In this paper, we relax the assumption on monotonicity of the dynamics to positive invariance of finitely generated max-plus cones. The properties of positively invariant sets play an important role in studying system-theoretic properties of dynamical systems \citep{Blanchini1999}. The positive invariance of polyhedral cones for positive switching systems has also been studied recently to generalise the Perron-Frobenius theory \citep{Forni2017}. 
	
	The main contribution of this paper with respect to the state-of-the-art is that we propose a general framework for studying stability of \Gls{SMPL} systems. Under this framework, we present different autonomous notions of stability and positive invariance for a general class of discrete-event systems. In addition, we analyse stability of \Gls{SMPL} systems under arbitrary and constrained switching sequences. 
	
	The paper is organised as follows. Section 2 gives background on the max-plus algebra. Section 3 presents a general framework for modelling discrete-event systems in the max-plus algebra. Section 4 presents the notions of stability and positive invariance for a general class of discrete-event systems in max-plus algebra. Section 5 then extends these notions to analyse stability of \Gls{SMPL} systems. The paper ends with concluding remarks in Section 6.
	\section{Preliminaries}

	The set of all positive integers up to $n$ is denoted as $\underline{n}=\{l\in\mathbb{N}\,|\,l\leq n\}$ where $\mathbb{N}=\{1,2,3,\dots\}$.
	
	The \textit{max-plus algebra}, $\mathbb{R}_{\mathrm{max}}=(\mathbb{R}_\varepsilon,\oplus,\otimes)$, consists of the set $\mathbb{R}_\varepsilon = \mathbb{R}\cup \{-\infty\}$ endowed with the max-plus addition $(a\oplus b = \mathrm{max}(a,b))$ and the max-plus multiplication $(a\otimes b = a+b)$ operations \citep{Baccelli1992}. The max-plus zero element is denoted as $\varepsilon = -\infty$ and the max-plus unit element as $\mathds{1}=0$. 
	 The max-plus vector and matrix operations can be defined analogously to the conventional algebra. The max-plus zero and identity matrices are denoted as $\mathcal{E}$ and $\mathcal{I}$ respectively. 
	 
	 The max-plus powers of a matrix are defined recursively as $\maxpow{A}{k+1} = \maxpow{A}{k}\otimes A$ for $k\in\mathbb{N}$ with $\maxpow{A}{1} = A$. For scalars $\gamma\in\mathbb{R}$, $c\in\mathbb{N}$, $\maxpow{\gamma}{c}=c\cdot\gamma$. The partial order $\leq$ is defined such that for vectors $x,y\in\mathbb{R}^n_\varepsilon$, $x\leq y\Leftrightarrow x\oplus y = y \Leftrightarrow x_i\leq y_i,\; \forall i\in\underline{n}$.  
	
	
	

	\textit{Norms.} The max-plus algebra is equipped with the $l^\infty$ norm defined as  \citep{heidergott2014max}:
	\begin{equation*}
	\begin{aligned}
	&\norm{x}_{\oplus,\infty} = \underset{i\in\underline{n}}{\mathrm{max}}\;x_i,\quad &&x\in \mathbb{R}_\varepsilon^n. 
\end{aligned}
	\end{equation*}
The Hilbert's projective norm in max-plus algebra is defined as \citep{heidergott2014max}
\begin{equation}\label{eq:0.2}
\begin{aligned}
\norm{x}_\mathbb{P} &= \underset{j\in \underline{n}}{\mathrm{max}}\;(x_j)-\underset{j\in \underline{n}}{\mathrm{min}}\;(x_j),\quad x\in\mathbb{R}^n\\
\norm{A}_\mathbb{P} &= \mathrm{max}\left\{\norm{[A]_{\cdot i}}_\mathbb{P}\;|\;i\in \underline{m}\right\}, \quad A\in\mathbb{R}^{n\times m}.
\end{aligned}
\end{equation}
We have $\norm{x+y}_\mathbb{P}\leq \norm{x}_\mathbb{P}+\norm{y}_\mathbb{P}$.

\textit{Function properties.} A function $g:\mathbb{R}^n_\varepsilon\to\mathbb{R}^n_\varepsilon$ is said to be \textit{homogeneous} if $g(\lambda + x) = \lambda + g(x)$, for all $\lambda\in\mathbb{R}$. The function $g$ is \textit{monotone} if for all $x,y\in\mathbb{R}^n_\varepsilon$, $x\leq y$ implies $g(x)\leq g(y)$. A monotone and homogeneous function is \textit{non-expansive} in the Hilbert's projective norm \citep{gunawardena2003max}, i.e.
\begin{equation}\label{eq:0.3}
\norm{g(x)-g(y)}_\mathbb{P}\leq \norm{x-y}_\mathbb{P},\quad \forall x,y\in\mathbb{R}^n.
\end{equation}
The non-expansiveness of the function $g(\cdot)$ implies that if  $g(x) = \lambda + x$, for some $x\in\mathbb{R}^n$ and $\lambda\in \mathbb{R}$, then all trajectories $\{g^k(x)\mid k\in\mathbb{N}\}$ are bounded in the Hilbert's projective norm \citep{Gaubert2004a}.
 
\textit{Max-plus eigenvalue problem.} The set of max-plus eigenvalues $\Lambda(A)$ of a matrix $A\in\mathbb{R}_\varepsilon^{n\times n}$ is defined as all the solutions to the following max-plus eigenvalue problem \citep{Cuninghame-Green1979a}:
\begin{equation}
\begin{aligned}
\exists z\in\mathbb{R}_\varepsilon^n,\;\;z\neq \mathcal{E}_{n\times 1}\;\;\text{s.t.}\;\;
A\otimes z = \lambda(A)\otimes z.
\end{aligned}
\end{equation}
Then $z$ is known as the max-plus eigenvector of $A$ corresponding to the max-plus eigenvalue $\lambda(A)$. The largest such eigenvalue of the matrix $A$ is denoted as $\overline{\lambda}(A)$. We also define $\lambda^*(A) = \min_{i\in\underline{n}}[A]_{ii}$ whenever the matrix $A$ has finite diagonal entries. A matrix normalised by a scalar $\mu\in\mathbb{R}$ is denoted as $[A_\mu]_{ij} = [A]_{ij}-\mu$. The matrix is irreducible if for every $i,j\in\underline{n}$ there exists a $k\in\mathbb{N}$ such that $[\maxpow{A}{k}]_{ij}\neq \varepsilon$.

The Kleene star of a matrix is defined as $A^\star = \mathcal{I}\oplus A\oplus \maxpow{A}{2}\oplus\maxpow{A}{3}\oplus\cdots$. It exists only when $\overline{\lambda}(A)\leq 0$. 

\textit{Max-plus cone.}
The analogue of a vector space in the max-plus semiring $\mathbb{R}_\epsilon$ is called a semimodule. For a matrix $V\in\mathbb{R}_\varepsilon^{n\times m}$, $\mathrm{span}_\oplus V$ denotes the max-plus semimodule of $\mathbb{R}_\varepsilon^n$ generated by the columns of $V$. A semimodule $\mathcal{V}\subset \mathbb{R}^n_\varepsilon$ is said to be \textit{finitely generated} if it can be expressed as $\mathcal{V}=\mathrm{span}_\oplus V$ for some finite integer $m$:
\begin{equation}
\mathcal{V} = \left\{\bigoplus_{i = 1}^{m}\alpha_i\otimes v_i\biggm\lvert \alpha_i\in\mathbb{R}_\varepsilon\right\}, \quad V=[v_1,\dots,v_m].
\end{equation}  
A finitely generated semimodule is called a \textit{max-plus cone} as it is closed under addition $\oplus$ of its elements and under multiplication $\otimes$ with scalars in $\mathbb{R}_\varepsilon$.

\textit{Eigenspaces.} Let $\alpha,\beta\in\mathbb{R}$ with $\beta\geq \alpha$. Then the slice space $S_\alpha^\beta(g)$ generated by a function $g(\cdot):\mathbb{R}_\varepsilon^n\to\mathbb{R}_\varepsilon^n$ is defined as \citep{Gaubert2004a}
\begin{equation}\label{eq:32}
\begin{aligned}
S_{\alpha}^{\beta}(g)&=\left\{x \in \mathbb{R}^{n} \,\mid\, \alpha+x \leq g(x) \leq \beta+x\right\}.
\end{aligned}
\end{equation}
It is also noted that for ${\alpha'}\leq \alpha$ and ${\beta'}\leq\beta$ with $\alpha'\leq \beta'$, we have $S_{{\alpha'}}^{{\beta'}}\subseteq S_\alpha^\beta$. If $g$ is monotone and homogeneous, the slice space is invariant with respect to $g$.

A slice space $S_\alpha^\beta$ of a function $g(\cdot)$ is said to be bounded in the Hilbert's projective norm if 
\begin{equation}\label{eq:0.4}
\exists \delta\in\mathbb{R}\; :\; \forall x\in S_{\alpha}^{\beta}(g),\; \norm{x}_\mathbb{P}\leq \delta.
\end{equation}
\section{Modelling discrete-event systems in the max-plus algebra}
We propose a general modelling framework for discrete-event systems in the max-plus algebra. Such models can incorporate the event-varying structure of the system along with the control algorithm. This serves as a basis for generating modelling classes for studying the stability properties of discrete-event systems in the max-plus algebra. Then we study \Gls{SMPL} systems as the open-loop case of this model.  

\subsection{General model in the max-plus algebra}
The dynamics of a discrete-event system with both continuous and discrete variables evolving over a discrete counter $k$ can be defined in max-plus algebra as
\begin{equation}\label{eq:23}
\begin{aligned}
x(k) &= f(x(k-1),l(k),u(k),r(k)), \;\; k\in\mathbb{N}\\
l(k) &= \phi(x(k-1), l(k-1), u(k),v(k),w(k)), \\
y(k) &= h(x(k),l(k),u(k),r(k)).
\end{aligned}
\end{equation}  
Here, the components $x_i(k)$  of the continuous state represent the time of $k$-th occurrence of the state event $i\in\underline{n}$. The system dynamics $f(\cdot)$ and $h(\cdot)$ represent the evolution of the continuous state and output, respectively.  The mode $l(\cdot)$ takes values in a finite discrete set. This allows modelling the discrete changes in the synchronisation and ordering structure of events. The dynamics of this discrete mode is specified by the function $\phi(\cdot)$. 

The continuous control input $u(\cdot)$ serves as a time delay to the autonomous occurrence of continuous states and output. The discrete control input $v(\cdot)$ influences the discrete state update dynamics via $\phi(\cdot)$. The continuous exogenous signal $r(\cdot)$ can either represent a reference signal and/or an uncertainty in the input $u(\cdot)$. The discrete exogenous signal $w(\cdot)$ can represent a scheduling signal and/or uncertainty in the discrete dynamics $\phi(\cdot)$. The model in the absence of continuous control and exogenous inputs is referred to as \textit{semi-autonomous.}

We assume that the system dynamics $f(\cdot)$ is \textit{not necessarily} monotone in the state vector $x$. This is due to the dependence of $f(\cdot)$ on the discrete state $l(\cdot)$ that could be generated by an exogenous signal $w(\cdot)$.   
\subsection{SMPL systems}\label{sec:SMPL}
The precedence constraints among the timing of event occurrences governed by synchronisation but no choice can be represented by the edges of a directed graph. A max-plus linear state space system represents the dynamics of this time evolution for a fixed directed graph. An \Gls{SMPL} system, on the other hand, can model event-varying graphs \citep{VanDenBoom2012b}. 

The dynamics of a non-autonomous \Gls{SMPL} system with $n_\mathrm{L}$ modes for event counter $k$ as an open-loop representation of the general discrete-event system in \eqref{eq:23} is defined as \citep{VandenBoom2006}:
\begin{equation}\label{eq:1}
\begin{aligned}
x(k) &= A^{(l(k))}\otimes x(k-1)\oplus B^{(l(k))}\otimes u(k),\;\; k\in \mathbb{N}\\
y(k) &= C^{(l(k))}\otimes x(k)\\
l(k) &= \phi(l(k-1),x(k-1),v(k),u(k),w(k))
\end{aligned}
\end{equation}
Here, the matrices $A^{(l)} \in\mathbb{R}_\varepsilon^{n\times n}$, $B^{(l)} \in\mathbb{R}_\varepsilon^{n\times m}$ and $C^{(l)} \in\mathbb{R}_\varepsilon^{n_\mathrm{y}\times n}$ are the system matrices for the $l(k)$-th mode. The function $\phi(\cdot)$ then specifies the switching rule.


In this paper, we present \textit{autonomous notions of stability} for \Gls{SMPL} system. This serves as a first step towards stability analysis of a general class of \Gls{SMPL} systems \eqref{eq:1} with continuous exogenous inputs. These notions pertain to the internal properties of the system as opposed to its input-output behaviour. Therefore, we restrict ourselves to the case when there is no max-plus additive input $u(k)$ or $r(k)$ to the system i.e., $B^{(l)}=\mathcal{E}_{n\times m}$ for all $l\in \underline{n_\mathrm{L}}$. Such systems are referred to as \textit{semi-autonomous}. 

	\section{Stability of discrete-event systems}
	In this section, we first recapitulate the different autonomous notions of stability for discrete-event systems. Then, we formulate these notions in the max-plus algebra. This allows for further analysis in the systems-theoretic framework. The distinctions in these notions are also established for a general class of discrete-event systems in the max-plus algebra. We also introduce the notion of positive invariance of certain max-plus cones to ascertain stability properties of such systems. This forms a basis for generalising stability notions to a general class of \Gls{SMPL} systems.
	\subsection{Introduction}
	The \textit{buffer level} in discrete-event systems is defined as the time delay between the occurrences of different events in either the same event cycle ($k$) or consecutive ones. The notion of stability is associated with the boundedness of the buffer levels \citep{Passino1998}. Asymptotic stability then implies that the buffer levels take constant values. This is only possible when the average growth rates of all the states become  equal to each other. 
	
	For a max-plus linear system, this average growth rate is given by the max-plus eigenvalue of the state matrix $A$ and is unique if the underlying graph is strongly connected \citep{Baccelli1992}. Moreover, there exists at least one finite max-plus eigenvector. Such max-plus eigenvectors specify a class of stable operation schedules. 
	
	A finite max-plus eigenvector represents a fixed point of the function $f(\cdot)$ \citep{gunawardena2003max}. There can exist multiple such max-plus eigenvectors corresponding to the same max-plus eigenvalue. When moving from a max-plus linear to an \Gls{SMPL} description, there can exist multiple max-plus eigenvalues and associated finite eigenvectors depending on the switching sequence. This makes the fixed-point theory for \Gls{SMPL} systems difficult to analyse. Moreover, the notions of bounded buffer levels for the events in the same event cycle and in consecutive ones are not equivalent any more. The following exposition presents a mathematical framework for understanding such differences. 
	We will first formulate autonomous notions of stability for a general discrete-event system in \eqref{eq:23}. Then we will impose \Gls{SMPL} dynamics on these notions.
	\subsection{Notions of stability}\label{sec:notion}
\begin{figure}
		\centering
		\begin{subfigure}[b]{0.425\textwidth}
			\includegraphics[width=\textwidth]{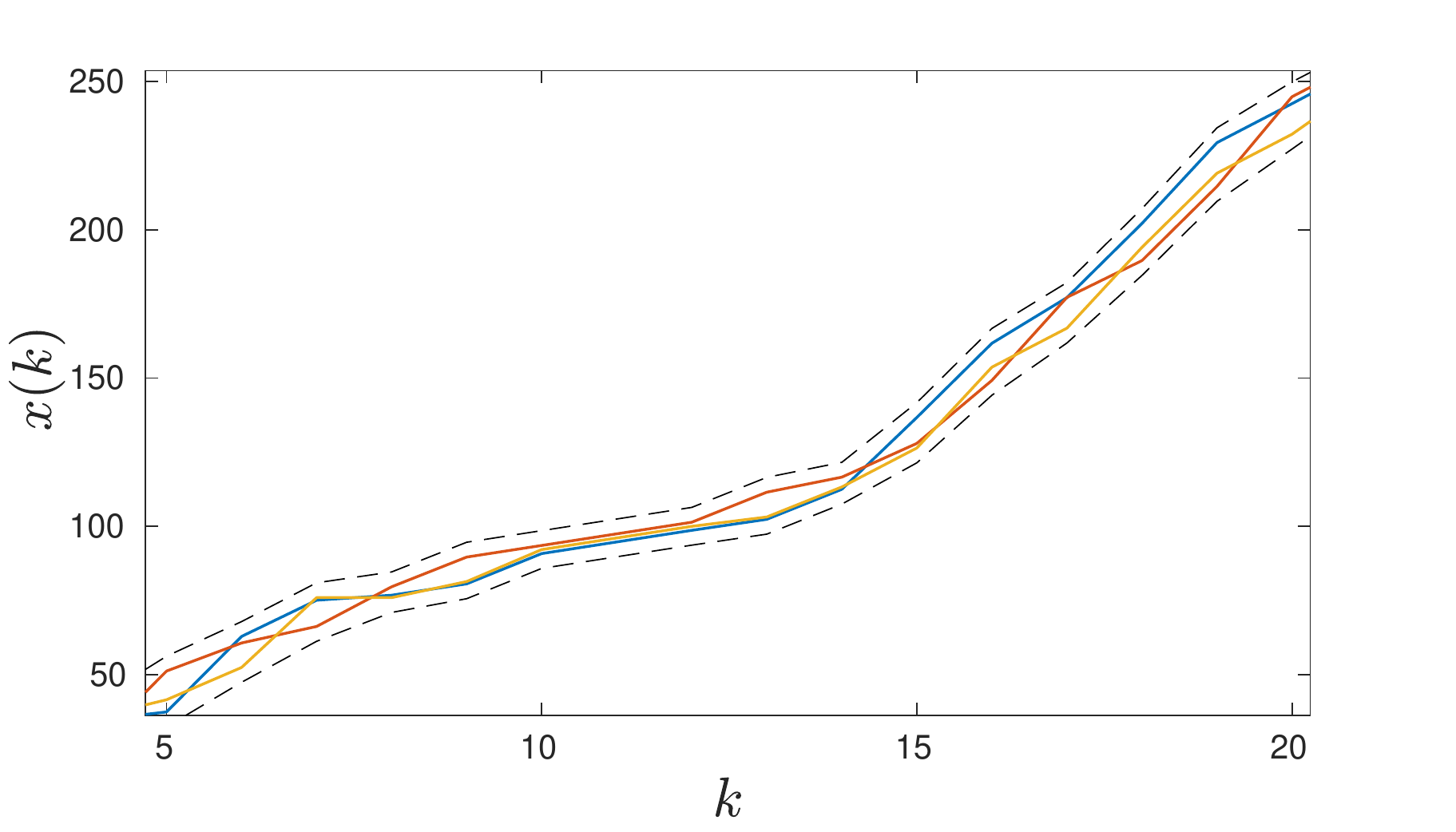}
			\caption{\label{fig:4a}}
		\end{subfigure}
		\begin{subfigure}[b]{0.425\textwidth}
			\includegraphics[width=\textwidth]{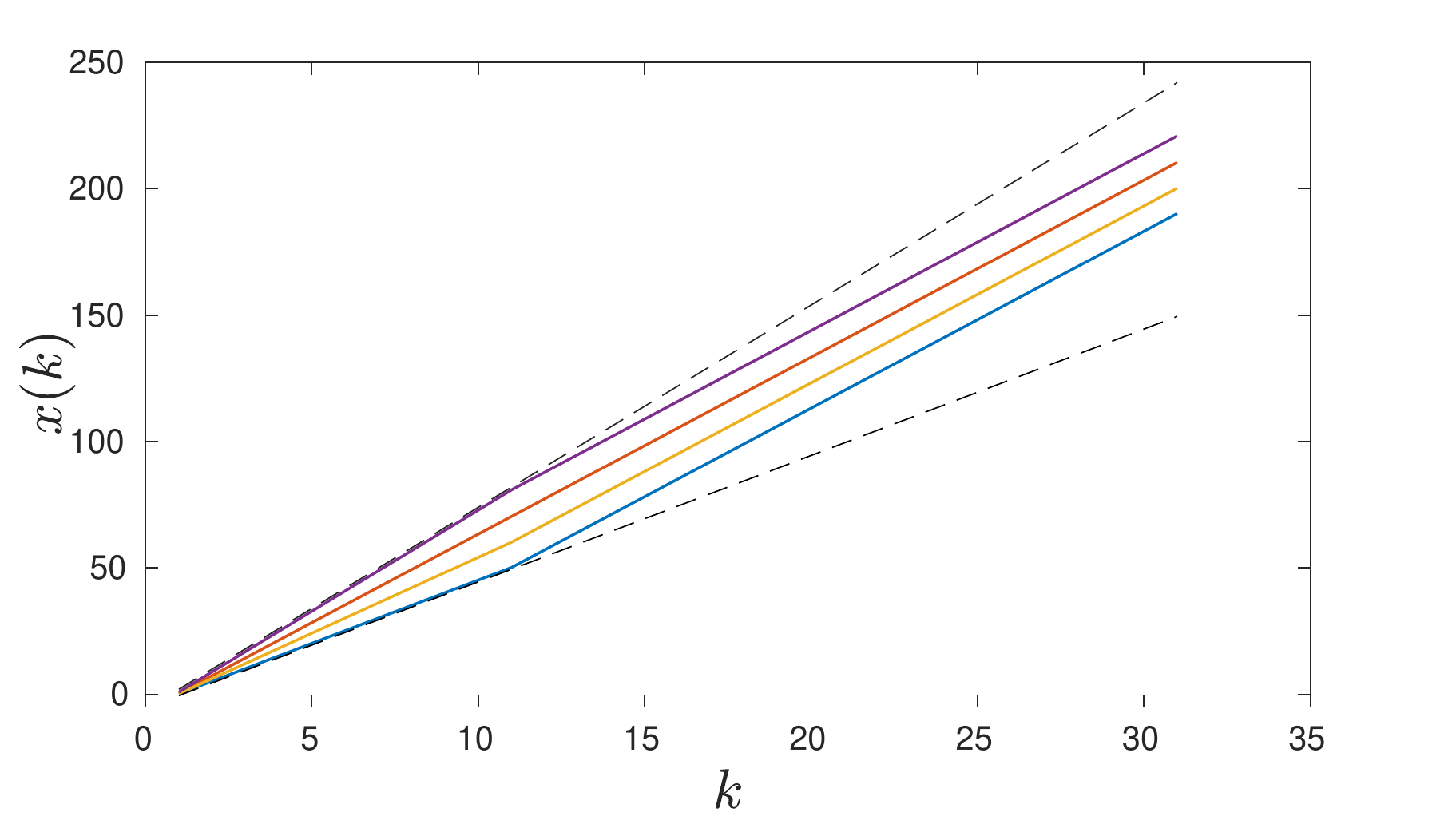}
			\caption{\label{fig:4b}}
		\end{subfigure}
	\caption{	\label{fig:4}Notions of stability: (a) Weakly bounded-buffer stability and (b) Max-plus Lipschitz stability. Dashed lines show the respective bounds on the trajectories.}
	\end{figure}

	In this paper, we focus only on the autonomous notions of stability for the system \eqref{eq:23} in absence of exogenous continuous inputs. The first set of notions deal with stability associated to boundedness of the buffer levels. The weak notion in this regard implies that the states are bounded from above and below (See Fig. \ref{fig:4a}). The stronger notion places component-wise bounds between the states. 
	
	\begin{definition}\label{def:1}
		A semi-autonomous discrete-event system is said to be \textit{max-plus weakly bounded-buffer stable} if for every $x_0\in\mathbb{R}_\varepsilon^n$, there exists a bound $M_\mathrm{x}(x_0)\in\mathbb{R}$  such that the states are bounded in the Hilbert's projective norm:
		\begin{equation}\label{eq:10}
		\begin{aligned}
		\norm{x(k;x_0)}_\mathbb{P} &\leq M_\mathrm{x}(x_0),\quad \forall k\in\mathbb{N}.
		\end{aligned}
		\end{equation}
	\end{definition}	 
	
	\textit{Asymptotic max-plus weakly bounded-buffer stability} is achieved if the norm in \eqref{eq:10} attains a constant value:
	\begin{equation}\label{eq:22}
	\underset{k\to +\infty}{\mathrm{lim}}\;\frac{1}{k}\,\norm{x(k;x_0)}_\mathbb{P}= 0. \vspace{-0.3cm}
	\end{equation}
	\hfill $\blacklozenge$
	
	For a max-plus linear system, the upper bound on $\norm{x}_\mathbb{P}$ can be obtained from the Hilbert's projective norm of the Kleene star of the system matrix normalised by its largest max-plus eigenvalue \cite[Theorem 3.104]{Baccelli1992}. The existence of this bound follows from the existence of a finite max-plus eigenvector of the system matrix \cite[Lemma 23-2]{Cuninghame-Green1979a}.
	
	The preceding notion is interesting when the required objective is to place
	bounds on the makespan of a discrete-event system in, for e.g., production networks.
	
	\begin{definition}\label{def:2}
		A semi-autonomous discrete-event system is said to be \textit{max-plus strongly bounded-buffer stable} if the states are bounded with respect to each other in the $l^\infty$ norm. Equivalently, the state trajectory is contained in a \textit{finitely generated} set $\mathcal{K}\subseteq\mathbb{R}^n$:
		\begin{equation}\label{eq:26}
\begin{aligned}
		&\exists Q\in\mathbb{R}^{n\times n},\quad [Q]_{ij} = q_{ij}\;\text{and}\\ &\exists\mathcal{K}=\left\{x \in \mathbb{R}^{n} \mid x_{i}-x_{j} \geq q_{ji}, \;\forall\, i, j \in \underline{n}\right\}\neq \emptyset
\end{aligned}
		\end{equation}
		such that 
		$
	x(k;x_0)\in\mathcal{K},\quad \forall k\in\mathbb{N}.
		$	 
{\textit{Asymptotic \MA strongly bounded-buffer stability}} is achieved if the component-wise bounds attain a constant value (compare with \eqref{eq:22}):  
$$\underset{k\to +\infty}{\mathrm{lim}}\;\frac{1}{k}\,\left(x_i(k)-x_j(k)\right)= 0,\quad \forall i,j\in\underline{n}. \hskip6.4em \blacklozenge$$ 
	\end{definition}
	The set $\mathcal{K}$ in \eqref{eq:26} can be represented as a max-plus cone:
	\begin{equation}\label{eq:13}
	\mathcal{K}=\left\{x \in \mathbb{R}^{n} \mid Q\otimes x \leq x\right\}.
	\end{equation}
	
	The max-plus cone $\mathcal{K}$ is finitely generated if $Q$ is an irreducible matrix \citep{Katz2007}. Moreover, $\mathcal{K}$ is an empty set if $Q$ has a max-plus eigenvalue strictly larger than $\mathds{1}$.  
	
	The stability in Definition \ref{def:2} can be studied as the invariance property of the system dynamics $f$ with respect to the max-plus cone $\mathcal{K}$. Such an invariance property for controlled max-plus linear systems $(A,B)$ has been studied by \cite{Katz2007} for matrices with integer entries. 
	
	Such a notion holds more significance when the time delays between the states are also constrained, for e.g., in railway networks. It can be proved using the properties of the Hilbert's projective norm that the strong notion implies the weaker notion but not vice versa. Explicitly, this means $\norm{Q}_\mathbb{P}\leq M_\mathrm{x}(x_0)$ for all $x_0\in\mathbb{R}^n$ in \eqref{eq:10}.

	The third notion deals with the boundedness of buffer levels associated to time delays in consecutive event cycles of a discrete-event system (See Fig. \ref{fig:4b}). This also implies a bound on the first-order difference of the state vector. 
	\begin{definition}\label{def:3}
		A semi-autonomous discrete-event system is said to be \textit{max-plus Lipschitz stable} if there exist upper and lower bounds ($\underline{M},\overline{M}\in\mathbb{R}$) on the first-order difference of the state trajectories: 
		\begin{equation}\label{eq:15}
		\underline{M}\leq\Delta\, x(k) = x(k)-x(k-1) \leq \overline{M}.
		\end{equation} 
	\end{definition}
\textit{Asymptotic max-plus Lipschitz stability} is achieved when the growth rate of state-trajectories become constant:

$
\forall i\in\underline{n},\;\exists\, c\in\mathbb{N}\;\; \text{s.t.}$
$$
\underset{k\to \infty}{\mathrm{lim}}\;\left[\Delta_c^2\, x(k) = x(k+c)-2x(k) + x(k-c)\right]_i = 0. \hskip2em \blacklozenge
$$
The stability notion in Definition \ref{def:3} does not, however, imply bounded-buffer stability (Definition \ref{def:1}). This can be observed in Fig. \ref{fig:4}. The initial trajectory in Fig. \ref{fig:4b} exhibits max-plus Lipschitz stability but not bounded-buffer stability. Moreover, a weakly bounded-buffer stable system might not exhibit constant bounds \eqref{eq:15} on the growth rate of the trajectories.    
	
	\begin{figure}
		\centering
		\begin{subfigure}[b]{0.433\textwidth}
			\includegraphics[width=\textwidth]{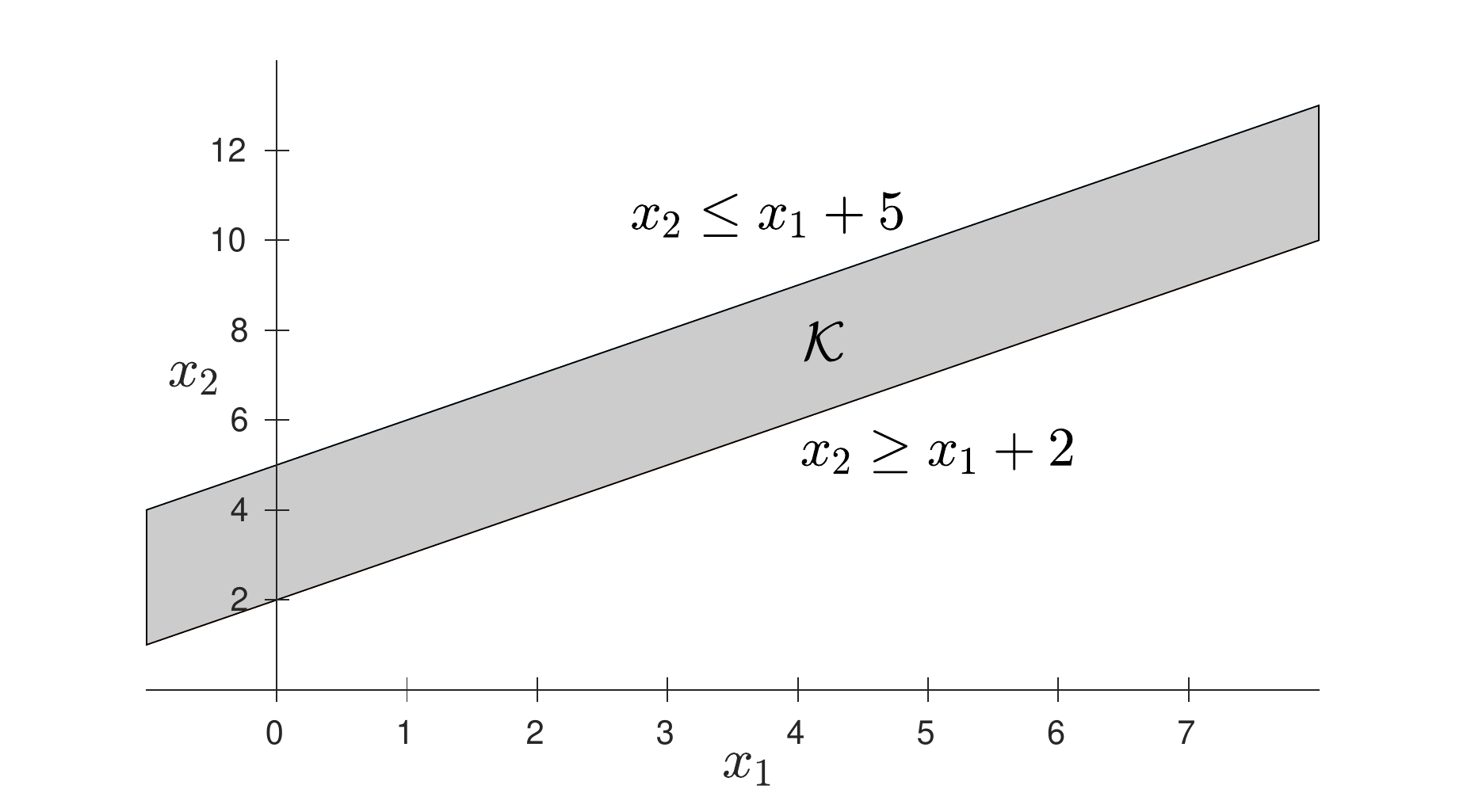}
			\caption{\label{fig:5a}}
		\end{subfigure}
		\begin{subfigure}[b]{0.433\textwidth}
			\includegraphics[width=\textwidth]{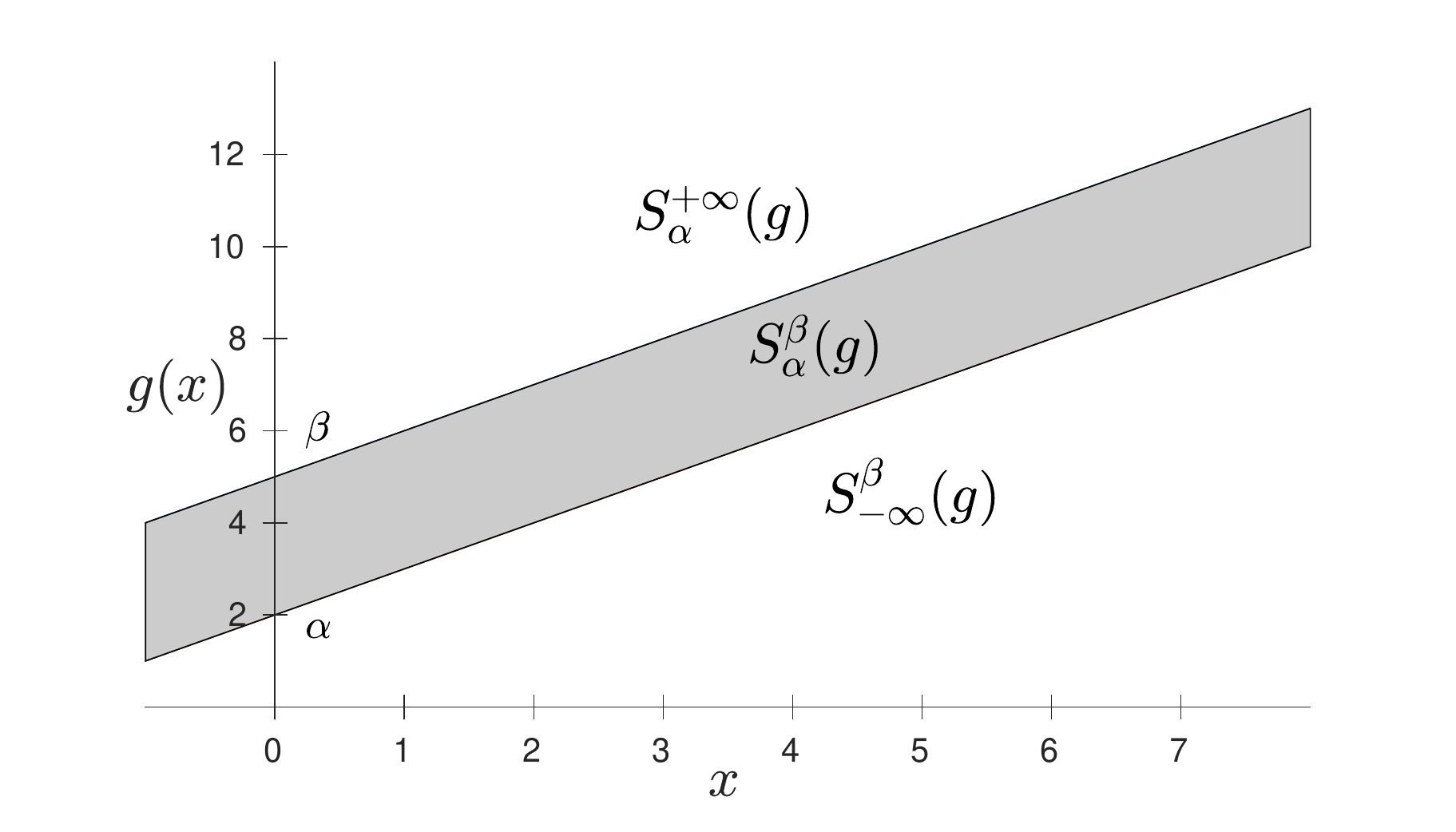}
			\caption{\label{fig:5b}}
		\end{subfigure}
		\caption{Max-plus cones: (a) Semimodule $\mathcal{K}=\left\{(x_1, x_2)^{\top} \in \mathbb{R}_{\varepsilon }^{2} \mid x_1+2 \leq x_2 \leq x_1 +5 \right\}$, and (b) Eigenspace $S^\beta_\alpha(g)=S^\beta_{-\infty}(g)\cap S^{+\infty}_\alpha(g)$ for $\alpha = 2$ and $\beta = 5$.}
	\end{figure}  
		\subsection{Positive invariance}\label{sec:cone}
	The dynamics of a semi-autonomous counterpart of the discrete-event system \eqref{eq:23} can be represented as $f(x,l)$. We say that a set $\mathcal{V}\subseteq\mathbb{R}^n_\varepsilon$ is positively invariant for the system dynamics if a state trajectory starting in this set remains in the set, $f(\mathcal{V},l)\subseteq \mathcal{V}$ for all possible $l\in\mathbb{N}$ \citep{Blanchini1999}. For convenience, the discrete state $l$ is left out of the notation whenever the statement holds for all values of $l$. 
	
	The first kind of positive invariant sets was discussed under the label max-plus strongly bounded-buffer stable systems in Definition \ref{def:2} (See Fig. \ref{fig:5a}). 
	
	The second kind of positive invariant sets are defined as slice spaces of functions in \eqref{eq:32}. These slice spaces represent max-plus cones when the generating function $g$ is monotone and homogeneous \citep{Gaubert2004a} (See Fig. \ref{fig:5b}).  
	
	We will show in Section \ref{sec:unistab} that the positive invariant set of first kind \eqref{eq:26} is a special case of the positive invariant set defined by slice spaces \eqref{eq:32}.  
	
	We study the Lipschitz stability of a system (Definition \ref{def:3}) as positive invariance of such slice spaces to dynamics. 
	\begin{proposition}\label{prop:1}
		Consider a monotone and homogeneous function $g:\mathbb{R}^n_\varepsilon\to\mathbb{R}^n_\varepsilon$ with a non-empty slice space $S_\alpha^\beta(g)\neq \emptyset$ for some $\alpha,\beta\in\mathbb{R}$ with $\beta\geq \alpha$ and with a finite max-plus eigenvector: $\exists x\in S_\alpha^\beta(g)$ such that $g(x) = \lambda \otimes x$ for some $\lambda\in[\alpha,\beta]$.
		
		Then, a semi-autonomous discrete-event system is max-plus Lipschitz stable if 		
		\begin{equation}\label{eq:17}
		f(S_\alpha^\beta(g))\subseteq S_\alpha^\beta(g)
		\end{equation}
		i.e., the slice $S_\alpha^\beta(g)$ is $f-$invariant.
		
		Moreover, the discrete-event system is max-plus weakly bounded buffer stable if  $S_\alpha^\beta(g)$ is bounded in the Hilbert's projective norm \eqref{eq:0.4}. 
	\end{proposition}
\textit{Proof:} As the function $g(\cdot)$ is monotone and homogeneous, the slice space $S_\alpha^\beta(g)$ is invariant with respect to $g(\cdot)$. Since the function $g(\cdot)$ is also non-expansive and has a positive eigenvector, trajectories $\{g^k(x)\mid k\in\mathbb{N}\}$ for all $x\in S_\alpha^\beta(g)$ are finite and bounded in the Hilbert's projective norm. Therefore, as the set $S_\alpha^\beta(g)$ is finitely generated, we have
	\begin{equation}\label{eq:14}
	\norm{g(x)-g(y)}_\mathbb{P} <+\infty,\quad \forall x,y\in S_\alpha^\beta(g).
	\end{equation} 	
From the positive invariance of $S_\alpha^\beta(g)$, we have
\begin{align}\label{eq:16}
\alpha + f(x)\leq g(f(x))\leq \beta+f(x).
\end{align}
Then, it follows that for all $x\in S_\alpha^\beta(g)$
\begin{equation}
\begin{aligned}
\norm{f(x)-g(x)}_\mathbb{P} &= \norm{f(x) - g(f(x))+g(f(x))-g(x)}_\mathbb{P}\\
&\leq \norm{f(x)-g(f(x))}_\mathbb{P} + \norm{g(f(x))-g(x)}_\mathbb{P}\\
&<+\infty
\end{aligned}
\end{equation}
where the last inequality follows from \eqref{eq:14} and \eqref{eq:16}.  
Then for all $x\in S_\alpha^\beta(g)$, we have
\begin{equation}
\begin{aligned}
\norm{f(x)-x}_\mathbb{P} &= \norm{f(x) - g(x)+g(x)-x}_\mathbb{P}\\
&\leq \norm{f(x)-g(x)}_\mathbb{P} + \norm{g(x)-x}_\mathbb{P}\\
&<+\infty.
\end{aligned}
\end{equation}
Therefore, for all $x\in S_\alpha^\beta(g)$ there exist $\alpha_x,\beta_x\in\mathbb{R}$ with $\beta_x\geq \alpha_x$  such that
$
\alpha_x\leq f(x)-x\leq \beta_x
$.
The smallest $\alpha_x$ and the largest $\beta_x$ over all $x\in S_\alpha^\beta(g)$ then provide lower ($\underline{M}$) and upper ($\overline{M}$) bounds for max-plus Lipschitz stability in \eqref{eq:15}. 

When the slice space $S_\alpha^\beta(g)$ is bounded in the projective norm, the bound on the slice space \eqref{eq:0.4} serves exactly as bound on the system trajectories \eqref{eq:10}. This results in max-plus weakly bounded-buffer stability. \hfill$\blacksquare$

We note that the boundedness \eqref{eq:0.4} of some slice space $S_\alpha^\beta$ of a monotone and homogeneous function $g$ implies the existence of a finite max-plus eigenvector of the function \citep{Gaubert2004a}. The converse implication might not hold. In the context of the preceding proposition, max-plus weakly bounded buffer stability implies max-plus Lipschitz stability but the converse is not necessarily true.
  
In essence, Proposition \ref{prop:1} relaxes the global monotonicity assumption on $f$ to a positive invariance property of an eigenspace generated by a monotone and homogenous function $g$. It is also noted here that it is generally difficult to prove boundedness of a slice space $S_\alpha^\beta(g)$ \citep{Gaubert2004a}. In this paper, we will restrict ourselves to max-plus linear functions $g$ for which results are available in the literature. 

	
	\section{Stability analysis for \Gls{SMPL} systems}\label{sec:PS}
	This section focuses on extending the stability notions provided in Section \ref{sec:notion} to semi-autonomous \Gls{SMPL} systems defined in Section \ref{sec:SMPL}. We present these results analogously to the stability results for switching systems in conventional algebra. These results are obtained by the means of positive invariance of the system dynamics with respect to the max-plus eigenspaces of certain classes of homogeneous and monotone functions.  
	
	\subsection{Stability of \Gls{SMPL} systems}
	We recall that the dynamics of a semi-autonomous switching systems \eqref{eq:1} can be described by a finite number of subsystems $f(x,l) = A^{(l)}\otimes x$, $l\in\underline{n_\mathrm{L}}$, along with a switching rule $\phi(\cdot)$. 
	
	When the sequence $l(k)$ is generated by an arbitrary switching rule, via $w(\cdot)$, we consider the scenario of \textit{uniform stability} of the \Gls{SMPL} system. This extension is parallel to the notion of \textit{uniform positivity} in positive switched systems as suggested by \cite{Forni2017}. Unlike the case of positive systems, it is however difficult to achieve contraction (mapping into the interior of a cone) in the Hilbert's projective metric \citep{Gaubert2004a}. Such a contraction, if it exists, would result in a common unique max-plus eigenvector of the matrices of the \Gls{SMPL} system, which is a very restrictive requirement.

	We also introduce the notion of \textit{path-complete stability} for the \Gls{SMPL} system analogously to the path-complete positivity notion for positive switched systems \citep{Forni2017}. This generalises the notions of uniform stability to \Gls{SMPL} systems with constrained switching sequences. This relies on the existence of multiple positively invariant sets. 
	
	
\subsection{Uniform stability}\label{sec:unistab}
We utilise the properties of positive invariance in a cone to give sufficient conditions of stability (defined in Section \ref{sec:notion}) for \Gls{SMPL} systems. A max plus matrix $Q \in\mathbb{R}_\varepsilon^{n\times n}$ serves as the most basic generating function for the max-plus eigenspace in Proposition \ref{prop:1}. To this end, we recall certain properties of such eigenspaces from the literature. 

The max-plus eigenspace generated by a max-plus matrix has long been a topic of investigation \citep{Butkovic2016}. The generating set of solutions for arbitrary $\alpha,\beta\in \mathbb{R}$ is usually studied for subeigenspaces $S_{-\infty}^\beta(Q)$ \citep{Sergeev2009a} and supereigenspaces $S_\alpha^{+\infty}(Q)$ \citep{Sergeev2015} separately. We study the slice spaces as intersection of these two eigenspaces. 

The subeigenspace  $S_{-\infty}^\beta(Q)$ is non-empty when $\beta\geq \overline{\lambda}(Q)$ and is generated as $S_{-\infty}^\beta(Q) = \mathrm{span}_\oplus(Q_\beta^\star)\cap\mathbb{R}^n$. When the matrix $Q$ is irreducible, we have that its Kleene star matrix $Q_\beta^\star$ has only finite elements. Therefore, $Q_\beta^\star$ and hence $S_{-\infty}^\beta(Q)$ are bounded in the projective norm.

There has also been a characterisation of the largest $\alpha$ values for which the supereigenspace $S_\alpha^{+\infty}$ is non-empty \citep{Butkovic2016}. We recall that $\lambda^*(Q) = \min_{i\in\underline{n}}[Q]_{ii}$ and $\lambda^*(Q)\leq \overline{\lambda}(Q)$. It is necessary to note that if $[Q_\alpha]_{ii}\geq 0$ for all $i\in\underline{n}$, then $S_\alpha^{+\infty}(Q)=\mathbb{R}^n$ \citep{Butkovic2016}.  

\begin{theorem}\label{thm:1}
	Consider a max-plus linear function $g(x) = Q\otimes x$, $Q\in\mathbb{R}_\varepsilon^{n\times n}$ with finite diagonal entries, with $S_\alpha^\beta(Q) \neq \emptyset$ for some $\alpha,\beta\in\mathbb{R}$ with $\beta\geq \alpha$. Moreover, it has a finite max-plus eigenvector: there exist $x\in S_\alpha^\beta(Q)$ and $\lambda\in[\alpha,\beta]$ such that $Q\otimes x = \lambda \otimes x$.
	
	Then, a semi-autonomous \Gls{SMPL} system is \textit{uniformly} max-plus Lipschitz stable if 
	\begin{equation}
	A^{(l)}S_\alpha^\beta(Q)\subseteq S_\alpha^\beta(Q),\quad \forall\,l\in\underline{n_\mathrm{L}}
	\end{equation}
	Moreover, if $Q$ is irreducible, the \Gls{SMPL} system is also \textit{uniformly} weakly bounded-buffer stable.
\end{theorem}
\textit{Proof:} It can be shown that the slice space $S_\alpha^\beta(Q)$ is finitely generated and non-empty for some $\alpha\leq \lambda^*(Q)$ and $\beta\geq \overline{\lambda}(Q)$. Moreover, $Q$ has a finite max-plus eigenvector in this slice space. Therefore, the first part follows from Proposition \ref{prop:1}.

Consider $\alpha\leq \lambda^*(Q)$, we have
\begin{equation}
\alpha\otimes x_i\leq [Q]_{ii}\otimes x_i\leq (Q\otimes x)_{i},\quad \forall i\in\underline{n},\;\; x\in\mathbb{R}^n.
\end{equation} 
Therefore, $S_\alpha^{+\infty}(Q) = \mathbb{R}^n$. We have $S_\alpha^\beta(Q) = S_{-\infty}^\beta(Q) = \mathrm{span}_\oplus (Q^\star_\beta)\cap \mathbb{R}^n$. Furthermore, if $Q$ is also irreducible and $Q_\beta$ has no max-plus eigenvalues strictly larger than  $\mathds{1}$, $Q_\beta^\star$ has only finite elements. This implies that the slice space $S_\alpha^\beta(Q)$ is bounded in the projective norm. Then, the system is also uniformly weakly bounded-buffer stable. 
\hfill
 $\blacksquare$

The preceding theorem proves the equivalence of Lipschitz stability and weakly bounded-buffer stability when the matrices of the \Gls{SMPL} system share a common slice space,
\begin{equation}\label{eq:25}
S_\alpha^\beta(Q)\subseteq \;S_\alpha^\beta(A^{(1)})\cap S_\alpha^\beta(A^{(2)})\cap\cdots\cap S_\alpha^\beta(A^{(n_\mathrm{L})})\neq \emptyset.
\end{equation}
Furthermore, the max-plus cone in \eqref{eq:13} is a special case of the max-plus cone presented in Section \ref{sec:cone} when the matrix $Q$ has only finite elements. If in addition, the system matrices $A^{(l)}$ max-plus commute with each other we obtain asymptotic strongly bounded-buffer stability of the \Gls{SMPL} system \citep{Katz2012}.    


\subsection{Path-complete stability}
The existence of a common positive invariant cone for all subsystems is a restrictive condition when the switching rule is not arbitrary. In this section, we consider the case of positive invariance when the generating function $g$ is composed of multiple matrices. This can be used, for example, to study invariance properties of \Gls{SMPL} systems when the switching function $\phi(\cdot)$ in \eqref{eq:1} models a finite-state automaton \citep{Forni2017}. The stability notion are then said to be \textit{path-complete} on the automaton.

\begin{theorem}
	Consider a set of $r\in\mathbb{N}$ matrices with finite diagonal entries:
	\begin{equation}\label{eq:21}
	\overline{Q} = \left\{Q^{(j)}\right\}_{j = 1}^{r}, \quad Q^{(j)}\in\mathbb{R}^{n\times n}_\varepsilon
	\end{equation}
	with $S_\alpha^\beta(Q^{(j)})\neq \emptyset$ for all $j\in\underline{r}$ and for some $\alpha,\beta\in\mathbb{R}$ with $\beta\geq \alpha$. Moreover, each matrix $Q^{(j)}\in\overline{Q}$ has a finite max-plus eigenvector: there exist $x^{(j)}\in S_\alpha^\beta(Q^{(j)})$ and $\lambda_j\in[\alpha,\beta]$ such that $Q^{(j)}\otimes x^{(j)} = \lambda_j\otimes x^{(j)}$.
	
	Then, a semi-autonomous \Gls{SMPL} system is \textit{path-complete} max-plus Lipschitz stable if there exist pairs $(i,j)\in\underline{r}^2$ such that 
	\begin{equation}\label{eq:24}
	A^{(l)}S_\alpha^\beta(Q^{(i)})\subseteq S_\alpha^\beta(Q^{(j)}),\quad \forall \,l\in\underline{n_\mathrm{L}}.
	\end{equation}
	
	Moreover, if all the matrices in the set $\overline{Q}$ are irreducible, the \Gls{SMPL} system is also \textit{path-complete} weakly bounded-buffer stable.
\end{theorem}
\textit{Proof:} We only give a proof for a special case of $g(\cdot)$: min of max-plus linear functions. The procedure can be extended to a more general class of constrained switching sequences. 

Consider the set defined  in \eqref{eq:21} and the following generating function of the invariant slice space \eqref{eq:17}: 
\begin{equation}\label{eq:18}
g(x) = \underset{j\in\underline{r}}{\mathrm{min}}\;[Q^{(j)}\otimes x],\quad x\in\mathbb{R}^n.
\end{equation}
Here, it is assumed that the minimum is element-wise and is attained for all $i\in\underline{n}$ with a single matrix $Q^{(j)}$, $j\in\underline{r}$. Such a function is monotone and homogeneous. 

Consider scalars $\alpha,\beta\in\mathbb{R}$ such that $\alpha\leq \lambda^*(Q^{(j)})$ and $\beta\geq \overline{\lambda}(Q^{(j)})$ for all $j\in\underline{r}$. Then the slice spaces $S_\alpha^\beta(Q^{(j)})$ are finitely generated and non-empty for all $j\in\underline{r}$. 

It is noted that the minimum in \eqref{eq:18} is attained for a single matrix $Q^{(j)}$. Hence, the slice space $S_\alpha^\beta(g)$ is finitely generated and non-empty. In addition, the existence of finite max-plus eigenvectors of all matrices $Q^{(j)}$, $j\in\underline{r}$, ensures the existence of a finite max-plus eigenvector of the function $g(\cdot)$ in \eqref{eq:18} that is again contained in $S_\alpha^\beta(g)$. 

For the positive invariance of the slice space $S_\alpha^\beta(g)$ to the system dynamics, we have
$
	A^{(l)}S_\alpha^\beta(g)\subseteq S_\alpha^\beta(g),\quad \forall\,l\in\underline{n_\mathrm{L}}.
$ 
Therefore, there exist pairs $(i,j)\in\underline{r}^2$ for every matrix $A^{(l)}$ such that \eqref{eq:24} is satisfied. This restricts the switching sequences allowed for the semi-autonomous \Gls{SMPL} system. Hence, the first part follows from Proposition \ref{prop:1}. 

Following the same arguments as in Theorem \ref{thm:1}, we have that every $S_\alpha^\beta(Q^{(j)})$ is finitely generated by the max-plus column span of its Kleene star. The boundedness of the slice space $S_\alpha^\beta(g)$ follows from the irreducibility of the generating matrices $Q^{(j)}$. Then the system is also \textit{path-complete} weakly bounded-buffer stable. 
\hfill
$\blacksquare$
 
The above theorem serves as a basis for studying stability of \Gls{SMPL} systems in case of constrained switching sequences. Unlike the notion of uniform stability, we do not require the existence of a common slice space \eqref{eq:25}. Moreover, relaxing the requirement of a common $\alpha$ and $\beta$ in the preceding theorem can allow us to provide tighter bounds on $\norm{x}_\mathbb{P}$ in \eqref{eq:10}. 

We finally note that the Proposition \ref{prop:1} provides a more general framework for stability analysis than provided in this section.
\subsection{Illustrations}
In this section, we present examples to illustrate the notions of uniform and path-complete stability for semi-autonomous \Gls{SMPL} systems. 

We consider the following semi-autonomous bimodal \Gls{SMPL} system \eqref{eq:1}:
\begin{equation}
A^{(1)} = \begin{bmatrix}
4 & \varepsilon\\1 &1
\end{bmatrix},\quad A^{(2)} = \begin{bmatrix}
3 & 3\\ \varepsilon& 6
\end{bmatrix}.
\end{equation}
In addition, we define a matrix $Q = A^{(1)}\oplus A^{(2)}$ as the max-plus summation of the system matrices. The slice space \eqref{eq:32} generated by the matrix $Q$ for $\alpha = 4$ and $\beta = 6$ can be obtained as $S_\alpha^\beta(Q) = \mathrm{span}_\oplus(Q_\beta^\star)\cap\mathbb{R}^n$,
\[S_{\alpha}^\beta(Q) = \left\{(x_1, x_2)^{\top} \in \mathbb{R}_{\varepsilon }^{2} \mid x_1-5 \leq x_2 \leq x_1 +3 \right\}.\]
It can then be evaluated that $A^{(l)}S_\alpha^\beta(Q)\subseteq S_\alpha^\beta(Q)$ for $l = 1,2$. This implies that the two matrices share an invariant max-plus cone. Moreover, the irreducibility of the matrix $Q$ implies that the \Gls{SMPL} system is also uniformly weakly bounded buffer stable under arbitrary switching.  
\begin{figure}
	\centering
		\includegraphics[width=0.15\textwidth]{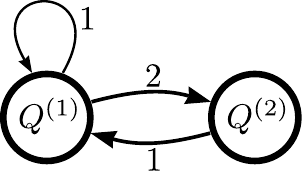}
	\caption{\label{fig:5}Automaton accepting constrained switching sequences corresponding to \eqref{eq:20}, \eqref{eq:27}. The edge labels represent mode $l$, the nodes represent max-plus cones.}
\end{figure}

We now consider another semi-autonomous bimodal \Gls{SMPL} system \eqref{eq:1}:
\begin{equation}\label{eq:20}
A^{(1)} = \begin{bmatrix}
4 & \varepsilon\\1 &1
\end{bmatrix},\quad A^{(2)} = \begin{bmatrix}
3 & \varepsilon\\ \varepsilon& 6
\end{bmatrix}.
\end{equation}
We define the following matrices:
\begin{equation}\label{eq:27}
Q^{(1)} =  \begin{bmatrix}
4 & 3\\1 & 1
\end{bmatrix},\quad Q^{(2)} = \begin{bmatrix}
4 & 0\\ 0& 4
\end{bmatrix}.
\end{equation}
The slice spaces generated by the matrices $Q^{(1)}$ and $Q^{(2)}$ with $\alpha = 0$ and $\beta = 4$ can be obtained as before,
\begin{equation}\label{eq:19}
\begin{aligned}
&S_{\alpha}^\beta(Q^{(1)}) = \left\{(x_1, x_2)^{\top} \in \mathbb{R}_{\varepsilon }^{2} | x_1-3 \leq x_2 \leq x_1 +1 \right\}\\
&S_{\alpha}^\beta(Q^{(2)}) = \left\{(x_1, x_2)^{\top} \in \mathbb{R}_{\varepsilon }^{2} | x_1-4 \leq x_2 \leq x_1 +4 \right\}
\end{aligned}
\end{equation}
It can then be evaluated that $A^{(l)}S_\alpha^\beta(Q^{(i)})\subseteq S_\alpha^\beta(Q^{(j)})$ for the triples $(l,i,j) \in \{(1,1,1),(2,1,2),(1,2,1)\}$ (Fig. \ref{fig:5}). Therefore, the system is path-complete weakly bounded buffer stable with respect to the max-plus cones \eqref{eq:19}.

\section{Conclusions and Future Directions}
In this paper, we have proposed a framework for studying stability of switching max-plus linear systems. We introduced autonomous notions of stability for a general class of discrete-event systems modelled in the max-plus algebra. We also relaxed the requirement of global monotonicity of the system dynamics to positive invariance of finitely generated max-plus cones. This allowed us to generalise the notions of stability to switching max-plus linear systems under arbitrary and constrained switching. Uniform stability is a restrictive condition when the switching sequence is not arbitrary. The notions of path-complete stability relax this requirement when the switching is constrained. 

In the future, we will look into the algorithmic aspects of the existence and computation of positively invariant max-plus cones. We also plan to extend this framework to non-autonomous notions of stability in the presence of inputs. Finally, we plan to extend the stability analysis procedure to closed-loop switching max-plus linear systems. 

\bibliography{library1}
	
\end{document}